\documentclass[10pt]{article}

\usepackage[]{amsmath,amscd,amsfonts,amsthm,amssymb}
\usepackage[greek, english]{babel} 
\languageattribute{greek}{polutoniko}

\usepackage{graphicx}
\usepackage[all]{xy}
\usepackage{psfrag}
\usepackage[lflt]{floatflt}
\newcommand{\bq}{\begin{quote}}
\newcommand{\eq}{\end{quote}}

\newcommand{\order}{\leqslant}

\newtheorem{Th}{Theorem}
\newtheorem{ax}{Axiom}
\newtheorem{lm}{Lemma}
\newtheorem{df}{Definition}
\newtheorem{pr}{Proposition}
\newtheorem{cl}{Corollary}
\newtheorem{re}{Remark}
\newtheorem{as}{Assumption}
\newtheorem{wg}{Wild Guess}
\newtheorem{ex}{Example}
\newcommand{\bth}{\begin{Th}\hspace{-5pt}{\bf .} \ }
\newcommand{\Eth}{\end{Th}}
\newcommand{\bax}{\begin{ax}\hspace{-5pt}{\bf .} \ }
\newcommand{\eax}{\end{ax}}
\newcommand{\blm}{\begin{lm}\hspace{-5pt}{\bf .} \ }
\newcommand{\elm}{\end{lm}}
\newcommand{\bdf}{\begin{df}\hspace{-5pt}{\bf .} \ }
\newcommand{\edf}{\end{df}}
\newcommand{\bpr}{\begin{pr}\hspace{-5pt}{\bf .} \ }
\newcommand{\epr}{\end{pr}}
\newcommand{\bcl}{\begin{cl}\hspace{-5pt}{\bf .} \ }
\newcommand{\ecl}{\end{cl}}
\newcommand{\bre}{\begin{re}\hspace{-5pt}{\bf .} \ }
\newcommand{\ere}{\end{re}}
\newcommand{\bas}{\begin{as}\hspace{-5pt}{\bf .} \ }
\newcommand{\eas}{\end{as}}
\newcommand{\bwg}{\begin{wg}\hspace{-5pt}{\bf .} \ }
\newcommand{\ewg}{\end{wg}}
\newcommand{\bex}{\begin{ex}\hspace{-5pt}{\bf .} \ }
\newcommand{\eex}{\end{ex}}

\newcommand{\bit}{\begin{itemize}}
\newcommand{\eit}{\end{itemize}\par\noindent}
\newcommand{\ben}{\begin{enumerate}}
\newcommand{\een}{\end{enumerate}\par\noindent}
\newcommand{\beq}{\begin{equation}}
\newcommand{\eeq}{\end{equation}}
\newcommand{\beqa}{\begin{eqnarray*}}
\newcommand{\eeqa}{\end{eqnarray*}\par\noindent}
\newcommand{\beqn}{\begin{eqnarray}}
\newcommand{\eeqn}{\end{eqnarray}\par\noindent}
\newcommand{\grk}{\selectlanguage{polutonikogreek}}
\newcommand{\eng}{\selectlanguage{english}} 
 
\usepackage{fancybox}

\title{\huge Newton vs. Leibniz: Intransparency vs. Inconsistency} 
\author{} 
\date{}

\begin{document}   

\maketitle  

\vspace{-1cm} 

%\centerline{\small \sc {[DRAFT]}}

\centerline{Karin Verelst}
\vspace{1mm}
\par
{\scriptsize
%\centerline{{\em FUND-CLEA}}
\vspace{1mm}
\centerline{{\em Vrije Universiteit Brussel\footnote{FUND-CLEA, Dept. of Mathematics.}}}

\centerline{{\em kverelst@vub.ac.be}}
} 

\bigskip
\bigskip
\bigskip

\begin{flushright}{\footnotesize {\em The only way to avoid becoming}\\{\em a metaphysician is to say nothing}\\{E.A. Burtt}}
\end{flushright}
\bigskip

\begin{abstract}

\frenchspacing \noindent \end{abstract} 

\subsection*{\sc Introduction}
%\bigskip

\frenchspacing \noindent The early modern debates on the nature of space enjoy a renewed interest in the recent philosophical literature. Focal points of the discussion are the opposition between the ``absolute'' vs. the ``relational '' conception of space and the related issue of its presumed substantival nature. Quintessential formulations of the key positions in the debate remain the viewpoints attributed to Newton and Leibniz. But scholarly disagreement concerning them remains stark. Especially with respect to Newton the defended interpretations range from anti-metaphysical readings (based on the notorious ``I do not feign hypotheses'') to at first glance almost theological readings that have him speculate about the relation between God and world. Leibniz, on the other hand, is presented as Newton's counterpart, who sets out tackling mechanical problems with an {\em a priori} metaphysical scheme --- inherited from Descartes --- already laid out in his mind. But Newton clearly conceived of himself as a natural philosopher \cite[p. 2]{JANa}, while Leibniz, certainly in his earlier years, would have insisted on his use of strictly mechanical concepts only \cite[Introduction]{FICH}. 

We believe that, in order to be able to judge on the ``metaphysical degree'' of the positions taken by our protagonists, we should understand more clearly what metaphysics itself is about. This question is often ignored. We shall show that all metaphysical theories share some crucial {\em structural} features, independently of their further conceptual commitments; they may even contradict each other. This structure shapes the procedures which govern the invention of ideas in the context of such theories, by codifying some onto-logical {\em a priori} assumptions regarding the consistency of reality into its bare conceptual framework. All these seemingly different theoretical approaches can be subsumed under a  general strategy developed to defeat the paradoxes which inevitably occur in our experience of the real. It consists of dividing the ``world'' into a ``substantial'' layer of identity and a ``phenomenal'' layer of change, connected by a relation of causality.
 But then again it remains unclear what is meant by ``substance'',  ``relation'' or even ``causation'' in these diverging contexts \cite{CASS}. In order to deal with these questions, we follow a historiographical line of approach, placing the early seventeenth century debates explicitly back within their own historical frame.  Many of the issues treated by our protagonists have much older roots and reach back into the past, even into Antiquity. Both Newton and Leibniz were very conscious of this fact. Only by taking it into account, we prepare ouselves for a more detailed study of the different ontological positions they took with respect to space and time, their ensuing different forms of ``mechanism'', as well as the conflicting methodologies they consequently developed. The specific strategies they each follow in order to avoid the paradoxes appear upon closer analysis as instances of the structural scheme mentioned above, where the causal connection bridging the ontological gap is rendered opaque rather than transparent in the case of Newton, while with Leibniz it is spelled out explicitly, but appears to be inconsistent, not a minor problem for someone to whom consistency is the criterion for existence {\em par excellence}.

The historical approach developed in this contribution might at first appear somewhat awkward to the more philosophically minded reader, because of the abundant use of source-material and the extensive discussion of secondary literature. But, in the kind of subject discussed here, subtleties of presumably only historical interest may turn out to have far reaching conceptual consequences which are lost to later times, at the cost of the repetition of debates that have long been dealt with before, more often than not to a degree of sophistication unmatched by whatever followed afterwards. The philosopher who wants to trace back these conceptual paths inevitably has to restore the complex fabric to which they belong in its entirety as much as he can, even if the task at hand is impossible right from the start. We can only hope for the indulgency of our readership with respect to this choice.

\subsection*{\sc Method and Paradox}

Since Antiquity it is known that theories claiming to decribe (an aspect of) the ``real world''  have to deal with paradoxical phenomena of plurality and motion. These paradoxes received their canonical form in the arguments of Zeno, and they have challenged ever since the descriptive as well as explanatory power of our theories about the world. But they seem hard, if not impossible, to overcome, except by what Poincar\'{e} so aptly calls ``un aveu d'impuissance'' \cite[p. 127]{POIa}: schemes to hide or avoid them, basically by the introduction of silent additional hypotheses like {\em a priori} conceptions of space and time, or refutations that turn out to be themselves circular, e.g. all part/whole arguments based on induction.\footnote{As pointed out by Poincar\'{e} in his chapter on ``Les Logiques Nouvelles'' \cite[pp. 141-147]{POIb}, Interesting comments on this topic also by K. Svozil, ``Physical Unknowables'' \cite{SVOa}.}  Another option is to assign to paradoxes explicitly a place in the description, in the hope that they will stay where they are and not swarm out like viruses until the body of knowledge collapses. This is the case of atomism (in the proper sense), infinitesimal calculus, paraconsistent logic and the like  \cite{PRIa}. Except maybe for the case of paraconsistent logic, all these approaches can be subsumed under a  general strategy developed to defeat paradoxes that dates back to Antiquity. It consist of dividing the ``world'' into a ``substantial'' layer of identity and a ``phenomenal'' layer of change, connected by a relation of causality. Each layer is characterised by a different kind of infinity, with its proper mode of generation: (simultaneous) division and (stepwise) addition [Arist., {\em Phys.} 204a 6].\footnote{For Aristotle's works we used the edition in the Loeb Classical Library \cite{LOEB}.} Aristotle called these the actual and the potential infinite. Throughout the middle ages they were used to contrast the infinity of God with what we humans can make of it: {\em infinitum simultaneum} vs. {\em infinitum successivum} \cite{MAI}.\footnote{I owe this reference to D. Strauss, in his paper on the  Excluded Middle \cite{STRAU}.} This strategy we shall henceforth call ``classical metaphysics'', and all theories exhibiting these structural features are ---  from our point of view --- to be considered as metaphysical. In Early Modernity its level of operation shifts from logic and metaphysics (in the traditional sense) to the foundations of mathematics and natural science. It remains visible in the  separate treatment of statics/cinematics (identity) and dynamics (change) in the new mechanical theories. Within this framework it will be possible to clarify and accurately distinguish the metaphysical differences between supposedly equivalent theories like Newtonian and Leibnizian (analytical) mechanics \cite{LANC}. The infinities pop up, closely linked to these metaphysical differences, in their formal approaches to infinitesimal calculus. Finally, it helps to expose where in both theories the original paradoxes hide, and how they are kept under control. In Newton's case they are relegated to an immovable  and omnipresent God who bridges the gap between actual and potential, between absolute and relative, between cause and effect by means of a law $F=ma$ which emanates directly from Him. We shall show that this relation between cause and effect justifies the enigmatic ``hypothesis I'' (in the second edition of the {\em Principia}), ``That the centre of the system of the world is immovable''. With Leibniz God grants perseverance and internal consistency to everything in an infinity of interconnected worlds, actual and potential, great and small. They are all modally interconnected due to Leibniz's principle of the best of possible worlds. On this basis Leibniz's principle of the equivalence of hypotheses, the idea that this frame of reference is as good as any other when describing the universe, rests. The cinematics that goes with it is evidently relativistic and its dynamics based on an encompassing principle of  conservation, so that there is no need to blur the notion of causality through the incomprehensible act of an omnipresent God. But Leibniz's system does not escape from incongruencies either, for his God, in order to keep an eye on all these worlds, has to be able to take on all these mutually interrelated, possible perspectives simultaneously.  It has been shown by Specker in a famous argument that this is inconsistent \cite{SPE}. Curiously enough, the Kochen-Specker theorem developed on the basis of Specker's original paper is a result related to the inevitability of certain inconsistencies in quantum theory \cite{KOCH}, so that one can ask what kind of structural link could exist between Leibnizian and quantum mechanics.

\subsection*{\sc Identity and Causality}

\frenchspacing \noindent In contemporary debates involving causality, there is no generally agreed upon, philosophically sound, univocal definition or even description of what causality is or implies. There are, moreover, good reasons to believe that the concept as it is used both in philosophy and in science is intrinsically pluriform.

\begin{quote} {\small \em What then is the answer to the title question?  We do not quite know. There is no single account of causation --- no theory of what causation {\em is} --- that is free of counterexamples. Nor is there any theory  of causation that tallies best with all our intuitions about what causes what. The persistent failure to find a fully adequate philosophical theory of causation may well make us sceptical about the prospects of such a theory. Perhaps, we are looking for unity where there is plurality --- for an analysis of a single concept, were there are many} \cite{PSI}.\end{quote}

\noindent Our present situation is thus strikingly similar to that of the seventeenth century philosophers engaged in the foundational debates of their time. After the destruction of the Peripatetic foundation for the then prevailing metaphysical edifice, Descartes's attempt to replace it had started a vivid debate in scholarly circles throughout Europe: {\em On the one hand, the notion of causality is central to the period's major positions and disputes in metaphysics and epistemology. On the other hand, few of the most prominent figures of the period enter into detailed or precise accounts of the relation of causal dependence or causal connection} \cite[p. 141]{WIL}. Causality loosely denotes a necessary relationship between one event --- the cause --- and another one --- the effect --- the latter one being somehow the direct consequence of the first, but the nature of the relationship between them often remains undisclosed, although it clearly implies some change occuring at least in the effect. But physical change implies time. Two things which are at least temporally separated are warranted to be ``the same'' in a non-trivial sense when they are causally interconnected. In the case of motion, they even are required to be identical except for the spatio-temporal parameters involved. This is impossible to verify, as none of the theoretical constructions granting causality can be {\em completely} reduced to the empirical level, as Hume famously pointed out. Conservation of identity is itself a far from innocent metaphysical assumption, as has been analysed in detail by \'{E}mile Meyerson in his {\em Identit\'{e} et R\'{e}alit\'{e}}, a book that merits far more attention than it usually receives.\footnote{{\em Ainsi le principe de causalit\'{e} n'est que le principe d'identit\'{e} appliqu\'{e} \`{a} {\bf \em l'existence} des objets {\bf \em dans le temps}} \cite[p. 38]{MEY}. The r\^{o}le of conserved quantities in causal theories has become something of a hype since Phil Dowe's paper \cite{DOW}. See M. Kistler for a criticism \cite{KIS}. None of these, however, mentions Meyerson's pioneering work. In what follows I refer to the English translation of Meyerson's book \cite{MEYa}.} Since it cannot be verified that the parts of the world --- the physical systems --- which are subject to this process are truly ``the same'', this has to be postulated in advance, and granted by means of some universal principle: {\em time-measurement depends, in the last analysis, on the existence of laws in nature (...) laws if they are to be knowable can only be so as a function of the changing of time} \cite[pp. 35, 32]{MEYa}. Thus, causal structures shed light on a theory's hidden metaphysical assumptions. But, if the common use of the concept of causality is to make sense at all, there should at least be some shared structural characteristics between all theories relying on it. 

In order to highlight the nature of these structures, let us start by taking a closer look at what we mean when we call something ``metaphysical''. A metaphysical system is not just a ``theory of what there is'' as Quine famously wants \cite{QUIN}. From its classical origins onwards, metaphysics not only tried to develop a ``picture of the world'', but as well to warrant the truth of the picture delivered; the problematic relation between the structure of the world and the structure of our discourse about it is from the very start at its heart. But we know at least since Plato that the connection provided by logic between ``knowledge'' and the ''real world'' is as much part of the problem as it is of its solution. Theories of everything inevitably amount into contradictions, and metaphysics was invented to do away with inconsistencies in the first place.\footnote{As Aristotle makes very clear in the first book of the {\em Metaphysics}. The idea to separate being from non-being both on the ontological and the epistemological level as a strategy to defeat the paradoxes and inconsistencies of his predecessors, is already developed by Plato in the {\em Theaetetus}, the {\em Sophist} and the {\em Statesman}. Indeed, Plato's philosophy marks the birth of metaphysics in every sense \cite{VERc}.} 

\begin{quote} {\small \em Two great warring traditions regarding consistency originated in the days of the Presocratics at the very dawn of philosophy. The one, going back to Heraclitus, insists that the world is not a consistent system and that, accordingly, coherent knowledge of it cannot be attained by man. ($\ldots$) The second tradition, going back to Parmenides, holds that the world is a  consistent system and that knowledge of it must correspondingly be coherent as  well, so that all contradictions must be eschewed.} \cite[introduction]{KIS}\end{quote}

\noindent Aristotle defines metaphysics as the theory of ``being {\em qua} \/being'' [\grk t`o >`on  ~<h >'on\eng ; Arist., {\em Met.} $\Gamma$, 1003a20], a theory about what it means or implies to ``be'' in its different --- existential and predicative --- senses, which certainly does make sense but nevertheless does not help us any further either. Indeed, ``being {\em qua} \/being'' is the most general statement possible {\em within}  the metaphysical scheme, because it already presupposes the principle of contradiction. Our claim now is that a metaphysical theory is a partial or total description of the world in which the following dual relationships hold:

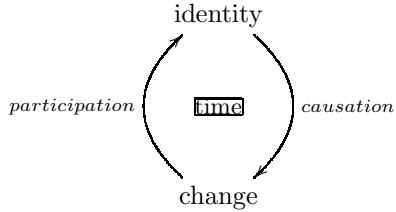
\[
\xymatrix{{\text{identity}} \ar@<1ex>@/^2pc/[dd]^{causation} \\*[F]{\text{\small{time}}}\\
\text{change}\ar@<1ex>@/^2pc/[uu]^{participation} }
\]

\noindent A {\bf metaphysical theory (CMet)} always involves a  general strategy which divides the ``world'' into a layer of stability (being) and one of change (non-being), {\em connected by a relation of causality}.  A crucial feature of the above scheme is that its backbone structure is logical in the classical sense: it is always grounded in some variant of the principle of contradiction (PC) [Arist., {\em Met.}, $\Gamma$,3,1005b(19-26); B 2,996b(30)], because it is designed to avoid the plurality (Parmenides) and motion (Heraclitus) paradoxes. This principle, as Aristotle very clearly states in the {\em Metaphysics}, therefore has to operate at the ontological and the epistemological level simultanuously, as do the paradoxes themselves \cite{LUK}. Furthermore, the arrows in the scheme are asymmetrical with respect to time; it is a dual, not an inverse relationship.\footnote{This scheme is more than just a fancy device. There is a recent branch of mathematics, category theory, which is particularly useful when dealing with this type of structural relationship. In category theory there is a specific kind of relation which captures behaviour and formal properties of structural connections between the local and the global level: {\bf adjunction}. Now our claim can be summarised as follows: causation in (meta)physical theories has the formal structure of an adjoint. Which adjoint? Let an example suffice for now to make the point: if one interprets causality in terms of order relationships, then we know already that this adjunction exists as the {\em Galois connection}. To put the idea a bit more formally: if causality can be expressed by $\order$, then there exists a categorical duality expressed by a pair of adjoint functors $Glob$ and $Loc$ (with $L \vdash G$) \cite[p. 96sq.]{BOR}. Galois connections abound in information theory, where $\order$ is interpreted as logically ``stronger than''(see \cite[p. 134]{VIC}), and in quantum logic, where it translates the idea of causal power into physically ``stronger than'' \cite[p. 20]{PIR}. We claim that appropriate adjoints exist for all possible interpretations of causation featuring the structural characteristics outlined above.} Indeed, causality as such by no means implies reversibility through time! If, in our reasonings about the real world, we take the classical connection between causal relations and logical entailment seriously --- and there is no reason we should not \cite{GHI} ---, then in our scheme causation and participation  coincide with respectively foward and backward implication, whereby the underlying ``effect'' is linked by a kind of necessity to its ``cause''. The level of identity thus connects instances which appear as distinguished facts on the empirical, particular level. In traditional metaphysics, this was accomplished by ``Forms'' of all stripes, e.g., Plato's {\em ideas}, or Aristotle's {\em substances}. Forms live in the realm of the immutable, the universal. they are part and parcel of the static part of the metaphysical set up of a theory. It should be observed that they are by no means material, but only serve to separate being from non-being, ``is'' from ``is-not''  both ontologically and epistemologically: they literally constitute the principle of identity \cite{VERc}. When traditional ontology crumbled at the verge of Early Modernity and became replaced, first by natural philosophy and eventually by modern science, the universalising r\^{o}le of substances had to be taken on by something entirely different: this is accomplished by the idea of {\em natural law} \cite{BRA}. Recall that the universal level in the metaphysical scheme serves to connect particular instances which are distinguished on the empirical level. In (experimental) science this is done by means of a universal statement, a {\em physical law} which connects measurable phenomena which are empirically very different, and expresses their relation quantitatively, in a mathematical way (basically, as a function of positions and velocities). This approach is an instance of what Burtt calls ``the central position of positivism itself'', the idea that it is possible to ``acquire truths about things without presupposing any theory of their ultimate nature; or, more simply, (...) to have a correct knowledge of the part without knowing the nature of the whole.'' \cite[p. 227-228]{BUR}  But here again the connection to the idea of causality remains problematic, because of the implicit introduction of the conservation of identity {\em through time}. \'{E}mile Meyerson analysed in his {\em Identit\'{e} et R\'{e}alit\'{e}} the precise relationship between identity and causality as follows:

\begin{quote}{\small \em The law states simply that, conditions happening to be modified in a determined manner, the actual properties of the substance must undergo an equally determined modification; whereas according to the causal principle there must be equality between cuases and effects Ñ that is, the original properties plus the change of conditions must equal the transformed properties.} \cite[p. 41]{MEYa}\end{quote}

\noindent That is to say, in order to go from lawful to causal behaviour, you need something more than the mere transition of states. You need something that warrants that $S$ at $t_1$ and $S$ at $t_2$ are still the {\em same} system. So the notion of ``natural law'' lies within the realm of the identical (the universal), and is thus part and parcel of the static part of the metaphysical set up of the theory. A physical law does not yet itself express causality. The question then arises what the nature is of the principle that shapes this invisible identity--conservation. Choices with respect to this are precisely what marks out the relevant metaphysical differences, even between {\em physical} theories: {\em (...) there is no escape from metaphysics, that is, from the final implications of any proposition or set of propositions. The only way to avoid becoming a metaphysician is to say nothing.} \cite[p. 227]{BUR}

To sum up, every CMet does come down to a more general global--local duality between the universal and the particular realm, which correspond to the layers of identity and change in the earlier scheme. In premodern theories, these layers are characterised by two kinds of infinity, the actual and the potential infinite, each with its own mode of generation: (simultaneous) division and (stepwise) addition  [Arist., {\em Phys.} 204a 6]. They inherit these infinities from the original infinities present in the paradoxes of Zeno.\footnote{That there are two different kinds of infinities involved in Zeno's arguments is plain, but rarely acknowledged. But see W.E. Abraham's paper on Zenonian Plurality \cite{ABR}. Also \cite{VERb}.} In what are traditionally called the plurality arguments, the Zenonian paradox appears as the coincidence of infinitely many large[s]-and-small[s] [\grk mikr`a te e~>inai ka`i m'egala \eng --- {\small DK} {\scriptsize{29B 1}}]\footnote{The extant fragments of the pre-Socratic philosophers are available in the critical edition by H. Diels and W. Kranz \cite{DK}. I follow scholarly custom in my references to that edition.}, the infinity of segments with and the infinity of segments without magnitude that result from Zeno's (simultaneously conceived) infinite division of a finite extension.\footnote{Remark that there is no dilemma involved (the text has `and' [\grk ka`i\eng], not `or' [\grk >'h? >'htoi\eng]). In fact this holds for the motion arguments as well. This was recognised by ancient commentators, e.g. Simplicius in his attestation that {\em [In his book, in which many arguments are put forward,] he [Zeno] shows {\bf \em in each} that stating a plurality comes down to stating a contradiction} [\grk kaj' <'ekaston de'iknusi, <'oti t~w poll`a e~>inai l'egonti sumb'anei tˆ >enant'ia l'egein \eng --- Simpl., {\em Phys.}, 139 (5) (cfr. {\small DK} {\scriptsize{29B 2}})]. I believe that on this analysis, it is possible to build a mathematically rigorous representation of all Zeno's paradoxes along the lines of Lawvere's categorical characterisation of ``cohesion'', as it implies a duality between the continuous and the discrete. We shall save this interesting subject for a future paper. Cfr. \cite{LAW}.} 

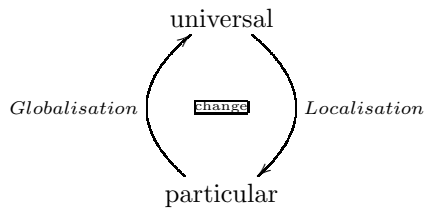
\[
\xymatrix{{\text{universal}} \ar@<1ex>@/^2pc/[dd]^{Localisation} \\*[F]{\text{\tiny{change}}}\\
\text{particular} \ar@<1ex>@/^2pc/[uu]^{Globalisation} }
\]

\noindent Plato developed his metaphysical system in an attempt to rebuke the contradictory arguments derived from the paradoxical conclusions reached by pre-Socratic philosophy on plurality and motion  as epitomised in the paradoxes of Zeno. He achieves this remarkable feat basically by transforming Zeno's paradoxical One-and-Many into a paraconsistent Large-and-Small.\footnote{See \cite{VERc}. The relevance of Platonic {\em diairesis} is stressed in Stenzel \cite{STE}.} Plato makes a clever shift from the physical to the mental, and considers every concept as an extension which can be divided --- Zenonian-wise --- into opposing halves, like Ôliving/non-livingÕ [Statesman, 261(b)], Ôfeathered/unfeatheredÕ [id., 266(e)] or Ôodd/evenÕ. This process ends when one bumps on undetermined parts or elements (\grk stoiqe~ia\eng) [Sophist, 252(b3)], that are not themselves capable anymore of specifying underlying parts [Statesman, 263(b)]. The number of steps needed to reach from the undetermined unity to this level of specification --- the proportion between part and whole --- then again defines the original concept, although this is not always possible [Philebus, 16(d)]. Note that the word ``specification'' makes sense here. Plato's system rests on a (paraconsistent) logic different from that of Aristotle, because it still comprises explicitly the two infinities both logically and ontologically \cite{VERc}. Aristotle proposed his alternative system in order to remedy certain criticisms on Plato's solution, concerning the status of infinity and the metaphysics of change. He literally puts a ``term'' (gr. \grk <'oros \eng; lat. {\em terminus}) to the ``largeness'' and ``smallness'' of an argument. In syllogistic theory these are the two extremes {\em maior} (large) and {\em minor} (small) connected to each other by the ``middle'': the maior is the premiss containing the predicate, the minor the subject of the conclusion, in which the middle term does not occur anymore. They effectively put limits to the Platonic infinities (the Large and the Small [\grk t`o m'ega ka`i t`o mikr'on\eng]) [cfr. {\em Phys.} 203a 15] that arise from logic conceived of as the division of an ``extended''  concept. From now on, what appears as a paradox in the real world manifests itself as an inconsistency in the realm of knowledge. 

One expects these embarassing infinities to reappear in the setting of natural philosophy and science, and this is evidently the case, as can be gauged from the Early Modern discussions on the foundations of calculus. \cite{MELa} These discussions, however, are not confined to the mathematical realm only: they arise precisely in the attempt to ``geometrise'' real motion. So Mariotte in his 1673 criticism of the ideas of Galilei: {\em mais ces raisonnements sont fond\'{e}s sur les divisions \`{a} l'infini, tant des vitesses que des espaces pass\'{e}s, et des temps de chutes, qui sont des raisonnements tr\`{e}s suspects (...)}.\footnote{Quoted by M. Blay in a delicious little book on the infinite \cite[p. 40]{BLA}.} There are other very explicit testimonia concerning the problematic nature of infinity, like the square number paradox Galilei put forward in his {\em Discorsi}. Descartes tries to settle this issue in his {\em Principia Philosophiae}, when he makes the important distinction between {\em indefinite} and {\em infinite} things:

\begin{quote} {\small \em And we shall call these things indefinite rather than infinite: first so that we may reserve the name of infinity for God alone, because in Him alone in every respect, not only do we recognize no limits, but also we understand positively that there are none; then too, because we do not in the same way understand other things in any respect to lack limits, but only negatively admit that their limits, if they have them, cannot be found by us.}
\footnote{{\em Haecque indefinita dicemus potius quam infinita: tum ut nomen infiniti soli Deo reservemus, quia in eo solo omni ex parte, non modo nullos limites agnoscimus, sed etiam positive nullos esse intelligimus; tum etiam, quia non eodem modo positive intelligimus alias res aliqua ex parte limitibus carere, sed negative tantum earum limites, si quos habeant, invenire a nobis non posse confitemur} [AT VIII-1, 18-25]. A discussion, relevant to our concerns, of this passage in Wilson \cite[p. 111]{WIL}. The translation is hers.}\end{quote}

\noindent This is nothing else than a variant of the Medieval distinction between {\em infinitum successivum} and {\em infinitum simultaneum} \cite{MAI}. So what started in Antiquity as an ontological part/whole dichotomy shifts with Descartes into a more mathematical direction, in harmony with the intellectual tendencies of the time \cite{BUR}. We shall soon encounter another pair of qualifications, relative and absolute, which will push the dichotomy to its metaphysical limits as it comes to embody the distinction between ontology and epistemology, between the lawful and the empirical, in fact as instances of a by then problematic mind/body relationship. 

Our claim up to now  is therefore that all CMet's structural requirements continue to hold true for the new theoretical edifices proper to XVII-th century natural philosophy, which can be read to a large extend as the attempt to answer the causal question along its main lines, but within a framework that does not allow anymore for the use of  key elements of ancient ontology as warrants for stability and identity, like, e.g., Aristotelian substance or Platonic ideas \cite{VERa}. Basically, the alternatives comprise some form of ``mechanism'', but this concept itself has a variety of meanings \cite[introduction, \S 2,]{DIJ}, \cite[p. 523 ft. 2]{MGUId}. We shall see that these can be reduced to the different underlying principles that warrant the conservation of identity shoring up any causal theory. With this in mind, we shall look in what follows at how Newton and Leibniz dealt with the question of causality and what the nature was of the solutions they proposed. 

\subsection*{\sc Substance as Substance and as Cause}

Before we can eventually turn to the XVII-th century, we have to take a closer look at some of the specific ontological commitments made by the founders of classical metaphysics, because their acceptance or rejection will play a crucial r\^{o}le in the early modern debates. A notion central to ancient versions of CMet is the notion of {\em substance}. In the common opinion, it is a kind of mysterious essence that defines somehow what a given thing {\em is} by ascribing certain properties to it. The thing ascribed to is called the subject, and its properties predicates; this correspondence between the verb and the fact `be' establishes the link between ancient ontology and epistemology: it grants that we can speak about something and at the same time know what we are speaking about. Substance is the key concept to ``the Aristotelian doctrine of `being-ness' ", and encompasses several important subtleties, of which we shall discuss some below. 

``Change'' in this context can mean two {\em different} things: 1) ``to begin to exist'', and 2) ``come to be such (come to have this form [\grk ti\eng])'' [Arist., {\em Phys.} 190b12-14]. The latter change takes place ``between the terms of an antithesis, such as `cultivated' and `uncultivated' '' \cite{COU}. Put like this it seems evident that the (ontological) subject [\grk <upoke'imenon\eng]  is the material stuff carrying those antithetical properties at given instances of time, and this is true to a certain extent, but it does not therefore imply that ``substance'' and ``matter'' are the same, even though {\em both} can function as the (logical) subject in a proposition. Aristotle on the contrary repeatedly stresses that matter is {\em not} substance. He adds that substance in the sense of a concrete individual with determinate characteristics consists of matter and form (\grk <'ulh + e'~idos\eng) [190b25-30], and that it changes on the level of its {\em eidos} or {\em ti} into certain antithetical properties, while staying the same as far as the material substratum is concerned.  Furthermore, Aristotle uses the expression \grk t'ode \eng [{\em Cat.} 5, 3b10] to refer to `this here', literally pointing out something, say Socrates, and the complex expression \grk t'ode ti \eng to refer to `this man here' (Socrates). The \grk t'o--de\eng: `this here', the bearer of existence, he calls primary substance.\footnote{cfr. Plato, [{\em Timaeus}, 49d-e]: ``(...) that in which they [the properties] each appear to keep coming to be and from which they subsequently perish, that is the only thing to refer to by means of the expressions `that' and `this' '' \cite{HEN}.} It is on this level that existential contradictions are completely ruled out  by the contradiction principle: one cannot be and not-be `here' at the same place, at the same time. ``Such'' [\grk ti\eng] is the secondary substance, the determined form (\grk o>us'ia) \eng [Phys. I, vii] or formal cause of `this'. `A man' in case the thing you are pointing at is Socrates. In order to refer to the essence {\em per se}, Aristotle uses the expression \grk ti >esti\eng: : `is such'. On this level, one does not deal anymore with existential contradictions, but with ``opposing terms'', contraries on the predicate-level which can very well coincide in an existing thing: something can be `man' and `not-man' (a god, say).  Further, non-essential, qualifications he calls \grk toi'onde\eng: so-and-so. They describe the how `this man' is, his inhering, accidental qualities, e.g., ``snubnosed'' in the case of Socrates [{\em Cat.} 5, 3b13-17]. In this latter case, contraries are evidently admissible as well. Clearly his distinctions between `subject' and `predicate' are different from and much less absolute  than ours to-day. And it is also clear that in the deictic context just outlined, it does not make sense to use (or merely think of) ``empty terms''! This is why existence is taken for granted in syllogistic arguments, and negation is always a secondary step in the development of a reasoning. It also explains why there is no need for Aristotle to explicitly formulate the principle of identity: when you speak deictically, in the here-and-now, in the present presence of a thing, identity is unproblematic.\footnote{\cite{BENa} {\em Position in space and time is a ``necessarily individuating property''. Determining such a  position involves an essential and ineliminable reference to another individual or position (...) 
To pick these out as the unique individuals or positions that they are we have to be able to {\bf \em relate them to ourselves or to the here-and-now}.} \cite[pp. 46-50]{QUI}} It is only when you move away from the present presence that you need {\em principles} to guarantee the truth of what you are saying, or, put differently: that your propositions do refer to something in reality. The zest of the whole argument is of course that ``this'' is universal because {\em you} are always present when you point or speak. Aristotle saw clearly that, for classical logic to be abstractly universal, i.e., outside the here-and-now, an extra principle to guarantee its validity was necessary. This is the principle of the excluded middle. Its ontological counterpart is the stability of the forms through time.\footnote{His key example are propositions with respect to the future, of which we do not yet know whether they will be true or false. For a discussion see, e.g., \cite{STRAU, FRE}. } 

It is the primary substance of which higher order --- more abstract --- substances (species, genera and other categorical qualifications) with a lesser degree of reality can be predicated. So the forms are immutable and matter as well, but for entirely different reasons! Material properties are conceived of as instances of substantial forms in an undifferentiated material substratum. Change from one state to the other is ontologically embedded in {\em the actualisation of potential properties} and governed by the four causes; the absolute endpoint of this sequence of actualisations is the fully actualised Prime and Unmoved Moving [Being], in relation to which everything else moves.\footnote{A principle, not a person! Aristotle's expression is \grk t`o pr~wton kino~un >ak'inhton \eng \cite[p. 368]{COUL}.} By introducing this divine metalevel, Aristotle succeeds in capturing motion while avoiding at the same time the existential contradictions, the {\em coincidentia oppositorum} that haunted all earlier attempts. This is important, because the need for an immutable substratum will have to be filled out differently once matter tranforms throughout the medieval school, not only into a bearer of properties by itself, but eventually also into the unique causal origin for change, in early modern atomism. We shall see that being immutable and being undifferentiated or universal end up by colliding in the XVII-th century into a new kind of meta-entity which serves as the personal warrant for individual exitence. This may account for, e.g., the awkward discussions in Locke on the difference between ``the essence of matter'', and ``the essences of material substances'', when he tries to come to terms with the impossibility to derive all secondary qualities from primary ones, and introduces the notion of ``superaddition'' --- an act of God --- to account for it.\footnote{An excellent discussion of the perplexities involved in M.D. Wilson, ``Superadded Properties; The Limits of Mechanism in Locke'', followed by a ``Reply to M. R. Ayers'', in {\em Mechanism}, especially p. 212.}

\subsection*{\sc Causality and Physical Law}

We saw that, in order to be able to implement the idea of causal connection into the body of a theory, the theory must comply to certain structural requirements. Since Antiquity, these structural requirements constitute the backbone for any consistent world description; that is why we simply use them to {\em define} a metaphysical  theory as such, independent of the further ontological commitments\footnote{This notion is used here in its general sense of the demands that a theory's truth impose on the world, cfr. \cite{ARA}.} that it implies. The ontological commitments are relevant, however, to identify the features in the theory which make the scheme work by granting the conservation of identity. Furthermore, we discussed before that the {\em universal level} in the CMet-scheme connects instances which are distinguished on the purely empirical, particular level of phenomenal reality. In experimental science this connection is encoded by a universal {\em physical law} which links phenomena which are empirically very different. The law expresses this relation quantitatively, ultimately as a function of positions and velocities. Now a minimal requirement with respect to the solidity of empirical verification as a means to check the reality of the supposed connection imposes itself: the {\em repeatability} of experimental results. Hence again the need for a universal ``scene'' on which the experiments can be established, but which remains itself  outside of observational reach:

\begin{quote} {\small \em  The general validity of the
principle that the universe presents the same aspect from every point (...) is accepted in modern physics as a necessary condition for the repeatability of experiments, since space and time are the only parameters which, at least in principle, are beyond the control of the experimenter and can not be reproduced at his will.} \cite[p. 84]{JAM}, \cite{VERa}\end{quote}

\noindent This introduces the notions of {\em space} and {\em time} into our discussion, as well as their problematic interrelation. The different ontological positions with respect to them, the ensuing different forms of ``mechanism'' and the common metaphysical core in the mechanical theories of Newton and Leibniz will be the major subject of the remainder of this article. What we now call classical mechanics ({\bf CMec}) in either form comprises a ``statical/cinematical'' and a ``dynamical'' level connected by a relation of causality. It continues to comply to our scheme and thus is a CMet. Indeed, by a simple amendment we could adjust Meyerson's quote cited above (cfr. ft. 5), so as to become applicable to mechanical theories specificially: {\em Ainsi le principe de causalit\'{e} n'est que le principe d'identit\'{e} appliqu\'{e} \`{a} l'existence des objets {\bf\em spatiaux} dans le temps}. It was once again \'{E}mile Meyerson who analysed for the first time the precise relationship between identity and causality as a relation between space and time: 

\begin{quote} {\em ...) the principle of causality demands the application to {\bf \em time} of a postulate, which, under the rule of lawfulness (l\'{e}galit\'{e}) alone, is only applied to {\bf \em space} (ne s'applique qu'\`{a} l'espace)} \cite[p. 32]{MEY}.\end{quote}

\noindent The notion of ``natural law'' therefore lies within the realm of the identical (the universal), and thus is part and parcel of the static part of the metaphysical set up of the theory. As we saw before, a physical law does not yet itself express causality.  One needs to bring together space and time close enough in order for the shift from causal to lawful to become possible. But this is problematic in itself, because of a fundamental ontological asymmetry between the two: {\em By nature, space and time are completely different, {\bf \em  each possible place in space is actual} in this moment, but for time {\bf \em only the present is actual}} \cite[p. 695]{PIRb}.This is a variant of the familiar problem that identity is unproblematic only in the present. The different strategies developed by the protagonists of XVII-th century natural philosophy to bridge this ontological gap will provide us with the key to unlock the doors that hide the metaphysical scaffold shoring up their respective theories.\\

{\large \bf Newton}\\

\noindent In recent years much research has been done on Newton's unpublished papers, providing us with a more realistic and quite different-than-usual picture of this most eminent among scientists. In fact, one of the main points to retain is that Newton was not, in our manner of speaking, a scientist, but --- as he would say himself --- a natural philosopher. Especially in the first part of his intellectually active life --- up to around 1700, thus including the first edition of the {\em Principia} --- he was hesitant concerning his methodological stance, and much more ready to take different but equally possible conceptual schemes into consideration when dealing with the fundamental philosophical and scientific problems of his time:

\begin{quote} {\em (..) to understand Newton's philosophy of science, one must not characterise his early and most creative periods of scientific thought by later slogans such as ``Hypotheses non fingo''. Rather we must see Newton's thought in its development as he progressed from a tolerance of certain types of hypotheses, especially speculations as to the cause of phenomena, to an alleged abhorrence of them all.} \cite[p. 163]{COH}\end{quote} 

\noindent A good place to start to understand Newton's approach is the text of a tract he wrote before venturing into the rigourously elaborated mathematical proofs of the {\em Principia}, the {\em De gravitatione et equipondio fluidorum}.\footnote{DG in what follows. An edition with translation is available in \cite{HAL}.} This text deals with a thorough criticism of the metaphysical positions elaborated by Descartes in his {\em Principia Philosophiae} \cite{DESa}. The ``general metaphysical position'' Newton develops in the DG provides the framework for his ideas on space, time, motion and existence for all his later work \cite{MGUIa}. The {\em De Gravitatione} played a pivotal r\^{o}le in the decisive metaphysical shift Newton made away from Cartesianism, I believe shortly before he set out to write his own {\em Principia}. It also bears witness to his taking the first steps towards what will become only much later, in the second and third edition of the {\em Principia}, his celebrated {\em Philosophia experimentalis} \cite{SHA}. Newton, while rejecting the positions of ``the Metaphysicians'' of his day, developed his own, original, metaphysical scheme, one of the goals of which was to do away, once and for all, with the need for ``hypotheses'' in what becomes, later on, science. The crucial step is taken by Newton in the ``Regulae'' at the beginning of Book III of the {\em Principia}, and lies in what Burtt calls ``the central position of positivism itself'', the idea that it is {\em possible to acquire truths about things without presupposing any theory of their ultimate nature.}\footnote{\cite[p. 227]{BUR} One should bear in mind that in the first edition of the Principia these ``regulae'' were still called ``hypotheses'', and had in important respects a different content. See Cohen's and Chaudhury's papers on this, discussed below.} This is connected to another seemingly innocent supposition, that there are isolisable systems in nature, a credo essential to any viable notion of scientific experimentation. In other words, by assimilating the metaphysics into the procedures, the {\em regulae}, the methods used for finding new truths about a given part of the world. {\em That a serious student of Newton fails to see that his master had a most important metaphysic, is an exceedingly interesting testimony to the pervading influence, throughout modern thought, of the Newtonian first philosophy} \cite[p. 229]{BUR}. Indeed, Newton's peers identified the {\em Principia} immediately as an attempt to replace the metaphysics of Descartes, whether they considered the attempt succesful or not. Even more, Newton did so himself, in anonymous tracts and in private communications! \cite[p. 3 ft. 6]{JANa}\footnote{Witness Huygens who refers to it as {\em (...) {\bf \em Newton dans ses Principes de Philosophie}, que je scay estre dans l'erreur (...)} in a letter to Leibniz concerning ``true motion'' in a discussion involving different notions of causality. \cite[OH X, n¡ 2854, p. 614 ]{HUY} (our bold). Huygens's reasons for this, by the editeurs of the {\em Oeuvres Compl\`{e}tes} (in ft 47) rightfully labelled, ``assertion remarquable'' will be the subject matter of a forthcoming paper.}
 
The intriguing fact is that Newton does not begin at all the DG as a criticism of Descartes's fundamental positions: {\em Newton clearly intended to write an elaborate treatise on hydrostatics; but, after completing a long criticism of Descartes, he seems to have lost interest in his original purpose} \cite[p. 76]{HAL}. On the contrary, the hydrostatics with which the DG begins seems to be intended as the development of a sound mathematical basis for the mechanical theory of motion based on the motion of fluids.\footnote{A point made at first by M.-F. Biarnais \cite[Introduction, pp. 9-10]{BIA}.} As late as 1682, Newton can write that the heavens consist of a material fluid that turns around in the direction of planetary motion: {\em materium coelorum fluidam esse. Materiam coelorum circa centrum systematis cosmici secundum cursam planetarum gyrare}\footnote{ULC Add. 3965.14, fl. 613r. This manuscript has been published in facsimile and transcribed with translation by J.A. Ruffner, as {\em Propositiones de Cometis}, \cite[pp. 260-263]{RUF}.}, a plain expression of the vortex theory of planetary motion \cite{AIT}. The reasons why he changed his mind have to do with his study of the comets in the early 1680s, and his changing point of view with respect to the r\^{o}le played by centrifugal motion in the explanation of gravity \cite{MELb}. This latter problem connects directly to the causal question. Both the Newtonians and the Mechanists saw circular motion (and its later generalisations to motions with a central force, like planetary orbits) as ``true'', i.e., as not merely apparent, and caused by real forces acting out there in the world. The difference between the two sides in the debate is about the nature of these forces.  Before 1681 Newton was convinced --- as were Descartes, Huygens and Leibniz --- that there is a close relation between inertia, centrifugal force and circular motion. A rotating body has a tendency to break away from its curvilinear path along the tangent in a point because of its inertia. This in turn generates centrifugal force, the tendency to recede from the centre of motion along the radius. Circular (elliptical) motion is the result of a dynamical equilibrium between the centrifugal force and gravity, understood as the pressure in the surrounding aetherial fluid caused by the effect of the mechanical forces exerted by the moving body on it; this is why centrifugal force (conceived of as a real force) is closely linked to the vortex theory.  It is only from 1684 onwards that Newton construes circular motion and its more general variants as caused by rectilinear inertia and gravity alone \cite[p. 33]{MELb}. 

Newton only comes to realise fully the untenability of Descartes's metaphysical positions when he tries to work them out in the {\em De Gravitatione} into a sound mathematical theory that is able to decribe the available facts; as we said before, this was his initial goal in the {\em De Gravitatione}, but he failed. While systematically scrutinising the positions defended by Descartes, Newton deepens his own insights, and succeeds in an increasingly clear formulation of his own alternative. The DG is the first place where we find him speaking about an immutable space which functions as the {\em principium individuationis} and as a global system of reference simultaneously, as opposed to the local one of the relative motion of bodies referring to themselves only, e.g., [DG,\cite[pp. 103-104]{HAL}].

Newton's solution is thus the rigourous parallellism between absolute space and absolute time, and the concommitant parallellism between relative spaces and times, whose finite individuality is granted for by the actual (always and everywhere) existence of their absolute counterparts.\footnote{The parallellism between space and time had been advanced at first by Gassendi, as pointed out by Bloch \cite[p. 179]{BLO}. For a discussion, see \cite{SMET}, and McGuire's discussion of this influence in his ``Existence''-paper \cite{MGUIa}. Newton knew Gassendi's works since his youth through a book by W. Charleton, and as an adult had several of them in his personal library. Traces of Newton's reading of Charleton are already found in his Trinity Notebook, dating from his student's days. It has been edited by J.E. McGuire and M..Tamny \cite[pp. 198-199]{MGUIb}.} This parallelism  requires infinite divisibility of both space and time, which Newton readily admits.\footnote{Another set of manuscripts, the {\em Tempus and Locus}-texts, relate to this as McGuire points out and elaborates in a comparative study of the different published and unpublished sources concerning this subject, which take up and refine the arguments of the DG \cite{MGUIc}.}

\begin{quote} {\em {\bf\em Space is an affection of being qua being.} No being exits or can exist which is not related to space in some way. God is {\bf\em everywhere}, created minds are {\bf\em somewhere}, and body is in the space that it occupies; and whatever is neither everywhere nor anywhere does not exist. (...) And the same may be asserted of duration: for certainly both are affections or attributes of a being according to which the quantity of any thing's existence is {\bf\em individuated} to the degree that the size of its presence and persistence is {\bf\em specified}.}\footnote{In Janiak's translation \cite[pp. 25-26]{JANb}. I slightly amended the first sentence.}\end{quote} 

\noindent The link to Newton's conception of he nature of body is not difficult to establish. After having clarified the nature of space and its relation to beings, he still has to explain what these beings --- material bodies --- themselves are, and what it means to say that they move. In the DG, Newton says --- still hesitatingly --- of ``beings which behave as if they were bodies'' that we can define them {\em as detemined quantities which omnipresent God endows with certain conditions. These conditions are: (1) that they be mobile (...) (2) (...) that they may be impenetrable (...) (3) that they can excite various perceptions of the senses (...)}. \cite[p. 28]{JANb} The omnipresent God clearly is the same as the one who individuates beings in infinite space. But it is only in the final drafts leading directly to the {\em Principia} that we find him resolve his problem by tying everything together in the notion of {\em inertia}. Empirical evidence shows that all accessible bodies gravitate towards the earth, also that the gravitational force decreases with an increase in spatial separation, and that it is proportional to the acceleration and the ``quantity of matter'' of the bodies involved. This new quantity Newton calls {\em mass}, and he defines it (def. 1) as ``arising from the volume and the density jointly.'' It ``cannot be intended and remitted'' (Rule III) and is therefore a universal property of matter, whence Newton can complete his list of ``essential qualities of bodies'' at the beginning of Book III \cite[pp. 94-95]{JANa}: 

\begin{quote} {\em The extension, hardness, impenetrability, mobility, and force of inertia of the whole arise from the extension, hardness, impenetrability, mobility, and force of inertia of each of the parts (...) And this is the basis of all natural philosophy... finally, if it is established universally by experiments (...) that all bodies (...) gravitate (...) and do so in proportion to the quantity of matter in each body (...) it will have to be concluded by this third rule that all bodies gravitate toward one another (...) Yet I am by no means affirming that gravity is essential to bodies. {\bf \em By inherent force I mean only the force of inertia}. This is immutable. Gravity is diminished as bodies recede from the earth.} [PR, Bk III, p. 796]\end{quote}

\noindent Inertia is a key concept to the extend that Newton starts his {\em Principia} with it, because it allows him to retain Descartes's law of inertia --- conservation of rectilinear motion in the absence of external forces --- freed from the burden of the latter's vortex-ontology.\footnote{Newton combined Descartes's idea of ``state'' with a transformed version of Keplerian inertia, associated with the concept of quantity of matter. Cohen discussed this on several occasions \cite[pp. 185-191]{COHb}; \cite{COHf}.} The actual infinity of absolute space fits in nicely as a means to make his own ontological construction work in accordance with his laws of motion, because infinite space provides the necessary framework within which inertial motion can be conceived in line with his first law --- as a test particle infinitely remote from whatever possible influence.\footnote{A clear description of the procedure to follow is in Lawden's book on tensor calculus: {\em (...) the evidence available suggets very strongly that if the motion {\bf\em  in a region infinitely remote from all other bodies} could be observed, then its motion would always prove to be uniform relative to our reference frame irrespective of the manner in which the motion was initiated. We shall accordingly regard the first law as asserting that, in a region of space remote from all other matter and empty save for a single test particle, a reference frame can be defined relative to which the particle will always have a uniform motion. Such a frame will be referred to as an {\em inertial frame}} \cite[p. 1]{LAWD}.} Space itself does not impede in any way the state of motion of the bodies it contains. Indeed, Newton goes as far as explicitly denying Descartes's objection with regard to the conceivability of infinite beings: 

\begin{quote} {\em If Descartes should now say that extension is not infinite but rather indefinite, he should be corrected by the grammarians. For the word `indefinite' ought never to be applied to that which actually is, but always looks to a future possibility (...) Thus matter is indefinitely divisible, but is always divided either finitely or infinitely (...) And so an indefinite space is one whose future magnitude is not yet determined; for indeed that which actually is, is not to be defined, but either does or does not have boundaries and so is either finite or infinite.} [DG \cite[p. 24]{JANb}.]\end{quote} 

\noindent This, again, explains why Newton can introduce the inertial mass of a system implicitly by definition, and not as an experimentally accessible quantity. Inasfar as the ``quantity of matter'' from the first definition of PR is an empirically accessible quantity, it is not inertial but gravitational mass, as is clear from Newton's mention of pendulums as a way of measuring it [prop. 6 PR Bk I]. In the third definition he introduces the ``inertia of matter'' as a {\em force (vis)}, the ``power of resisting by which every body (...) perseveres in its state of resting or moving''. Given his First Law, he can thus allow himself to speak in the Second Law only of forces as well. But, while inertial and gravitational mass are experimentally proportional and under a good choice of units even equal, they are by no means the same quantity. Newton's theory does not provide us with any explanation of this remarkable coincidence.\footnote{The complexities of this are discussed by Cohen \cite{COHb}.} It will help us, however, to uncover eventually where the CMet-scaffold that supports also this theory hides, once we connect it to a seemingly bizarre ``hypothesis'' in Book III that survived all revisions of the {\em Principia}. Let us allow ourselves a small anachronism and take the usual (differential) formulation of the second law to see how and where the concept of mass it uses differs from that appearing in Newton's celebrated law of universal gravitation. The Second Law in Newton's own words:

\begin{quote}{\em Mutationem motus proportionalem esse vi motrici impressae, \& fieri secundam lineam rectam qua vis illa imprimitur [A change in motion is proportional to the motive force impressed and takes place along the straight line in which that force is impressed].}\footnote{PR, Bk I, Var., p. 54 (16-17). Translation from I.B. Cohen and A. Whitman \cite[p. 416]{COHc}.}\end{quote}  

\noindent then becomes

 \beq F = ma\eeq 
 
 \noindent For the determination of the masses of celestial bodies, a different principle is used, the law of gravitation:\\

\beq F_g = G\, \frac {m\, m'}{r^2}\eeq

\noindent What one measures here is not mass as proportional between acceleration and force, but mass as an expression of the {\em strength of attraction between bodies}\footnote{A detailed analysis in H. Poincar\'{e}, ``La m\'{e}chanique classique'', in \cite[pp. 112-129]{POIa}.}, not the inertial, but the gravitational mass.  The proportionality constant $G$ in this case is the universal gravitation constant. Now we can evidently take one of the two bodies in equation (2) as the reference mass in whose gravitational field the other body moves. The equation than becomes:

\beq F_g = g m'\eeq

\noindent with  $g = G\,m/r^2$; $g$ is the strength of gravitational field of body $m$. Despite their formal similarities, eqs. (1) and (3) are not the same, since the mass in the former is inertial, in the latter gravitational. As said before, Newton's theory does not provide us with a reason nor an explanation for their identity. 

If the attraction were inversely proportional to the square of the distance alone, then it would be possible to determine the masses of the bodies involved by measuring their relative motions by means of the equation 

\[f = \frac{1}{r^2}\quad ;\]

\noindent but this is not the case. That the proportionality constant is a {\em constant} in Newton's Second Law is necessary for the consistency of the system as a whole: it grants the identity of bodies under the influence of external forces, like in rotational displacement (isotropy of space).  Now we want to postulate that these two different masses are nevertheless equal, because otherwise it is impossible to define consistently accelerated motion as the interaction between gravity and rectilinear motion alone (homogeneity of space).\footnote{This difference implicitly codifies the different status rotation and translation have in Newtonian Mechanics. Cfr. Meyerson on this crucial this point: {\em (...) our belief in the homogeneity of space implies something more than the persistence of laws. We are, indeed, convinced that not only laws Ñ that is, the relarions between things Ñ but even things themselves are not modified by their displacement in space.} \cite[p. 37]{MEYa}.} Only we have no means to rigourously verify this, and rigour is of the essence here, as Newton himself admits.\footnote{{\em Thus, instead of absolute places and motions we use relative ones, which is not inappropriate in ordinary human affairs, although in philosophy abstraction form the senses is required. For it is possible that there is no body truly at rest to which places and motions may be referred.} PR, Bk I (Scholium to the Definitions), in the translation by Cohen and Whitman \cite[p. 411]{COHc}.}  So in order to warrant this condition another principle is needed. This is Newton's Third Law. If one accepts the Third Law (equality of action and reaction) as an axiom, one can establish the constancy of this proportionality of the forces two bodies $A$ and $B$ are exerting on each other. But then again, this would be perfect only in the case the two bodies were alone in the world, so as to not be influenced by anything else, exactly as in the case of defining inertial motion. But in the real world this is never the case; one should, theoretically at least, decompose the acceleration of a body $A$ and determine which component is the consequence of the action of $B$. The validity of this procedure depends on our accepting the simple mutual addition of forces, which again imposes other auxiliary hypotheses, namely that the action of the bodies involved works along the straight line that connects their centres of gravity and depends only on their mutual distance, i.e., that actions can be reduced to forces exerted by mass points, in other words, they are central forces. This idea evidently can be generalised:

\begin{quote} {\em Corollary 4. The common centre of gravity of two or more bodies does not change its state whether of motion or of rest as a result of the actions of the bodies upon one another, and therefore the common centre of gravity of all bodies acting upon one another (...) either is at rest or moves uniformly straight forward.} [\cite[p. 421]{COHc}]\end{quote} 

\noindent Let us recapitulate. In Book I, Newton introduces absolute space (and time) firstly to account for inertial mass, and thus for the at least theoretically rigourous distinction between absolute (true)' and relative (apparent) motion, secondly as the immutable background scene against which thrustworthy empirical measurements of relative quantities are possible. It is safe to assume that all systems dealt with in book I are finite and thus relative, and all empirical reasonings concerning them do not need absolute space as such. Even if the systems considered are large, one can do with approximative inertial frames with respect to their centres of gravity, like the fixed stars relative to the sun as the centre of the solar system. But then again, the Third Law is a {\em law}: it is universally valid, so even the law of the motion of the centre of gravity is rigourously true only when applied to the universe as a whole. But this implies that, in order to find the (theoretical) values of the gravitational masses, you need to know the speed of the gravitational centre of the universe as a whole, which obviously is impossible, since we can --- fully in agreement with Newton's own point of view --- only measure relative motions. 

\begin{quote} {\em  But no system exists which is abstracted from all external
action; every part of the universe is subject, more or less, to the action of the other parts.  {\bf \em The law of the motion of the centre of gravity is only rigorously true when applied to the whole universe.} But then, to obtain the values of the masses we must find the motion of the centre of gravity of the universe. The absurdity of this conclusion is obvious; the motion of the centre of gravity of the universe will be for ever to us unknown. \cite[p. 57]{POIc}.} (Italics in the original.)\end{quote}

\noindent Contemporary scientists escape by saying that the masses are merely coefficients needed to execute certain calculations \cite[p. 126-127]{POIa}, but this was obviously not Newton's idea of them. From Newton's point of view, it is indispensible for the logical consistency of the theory to know the cinematical state of the centre of the universe in order to know theoretically what the true values for the masses are. Homogeneity and isotropy of space --- physically expressed in the equivalence of inertial and gravitational mass \cite{COHd} --- has to be supplemented by something that ultimately grants the applicability of this principle of identity universally, exactly as required by Meyerson's causal criterion. Here we find the reason for Newton's introduction of a remarkable Hypothesis I in Book III of the {\em Principia}, a hypothesis which has stirred the peace in the literature concerning it because it seems so strangely at odds with corollary 4 in Book I.\footnote{To the astonishment of several eminent commentators, like Hermann Weyl \cite[p. 71]{WEY}. See on this also Cohen \cite{COH}.} In the Third Book of the {\em Principia}\footnote{Originally planned to be the second book of the {\em Principia}, but which became the third after Newton inserted a new second book in which he gave his final treatment of the original problem dealt with in the {\em De Gravitatione}, fluid mechanics. He drafted it in early 1685, but this draft was only published posthumously as Newton's {\em System of the World} \cite[p. 262]{RUF}. For an edition and an explanation of the title, see \cite[p. xi]{COHe}. More on Book II PR in \cite{SMI}}, {\em De Mundi Systemate}, of which we have several preceding drafts, we find a hypothesis that survived all subsequent revisions and the transition from the hypotheses in PR I to rules in PR II \& III. This hypothesis continues to stand out as a hypothesis because it is a hypothesis in the proper sense of the word \cite{CHAU}. {\em And what a hypothesis this HYPOTHESIS I proves to be} \cite[pp. 165-166]{COH}:
\bigskip

\centerline{HYPOTHESIS I}

\centerline{\em Centrum systematis mundani quiescere}
\centerline{\em [That the centre of the system of the world is immovable].}

\bigskip

\noindent Newton adds: ``This is acknowledged by all, while some contend that the earth, others that the sun, is fixed in that centre. Let us see what may from hence follow.'' And he adds as a consequence:

\begin{quote} {\em Commune centrum gravitatis terrae, solis \& planetarum omnium quiescere [That the common centre of gravity of the earth, the sun, and all the planets, is immovable].}\end{quote}

\noindent This provides the theoretical, not just an approximative, reason for the fact that it is acceptable to use the sun as an inertial frame compared to the fixed stars with respect to which all other frames can be defined. To put it a bit boldly: in absolute space, all real motions are relative with respect to it. Newtonian mechanics is absolutely relative.\footnote{Cfr. Janiak's interpretation: {\em (...) we ought to distinguish absolute from relative space and time in order to understand true motion as a change of absolute place over time. (...) This move also enables Newton to save the perceptible effects of accelerating bodies --- most famously noted in the examples of the rotating bucket and the connected globes in the Scholium --- since all accelerations can be understood as true motions within absolute space.} \cite[p. 50]{JANa}.} In this sense --- and in this sense alone --- it is possible to appreciate fully the zest of Newton's Bucket argument as a precursor to Mach's principle.\footnote{This is confirmed rather than refuted by Barbour's argument on the ``Universe at large'', even if he believes it holds for Leibnizian mechanics only: {\em a fully relational (and hence Machain) theory should start by considering the relative motion of the universe treated as a single entity and then recover the motion of subsystems within the background provided by the Universe at large.} \cite{BARB} I believe that Barbour misses a crucial point, however, which prevents him to see that in this respect Newtonian and Leibnizian mechanics ultimately agree. The reason is --- again --- that he takes only finite, or at best indefinite (countably infinite) ``universes'' into account. We shall come back to this in the chapter on Leibniz, below.} Absolute space is necessarily prior to inertia and thus to quantity of motion, not the other way around, when the whole of the system of the world is at stake. Therefore we have to disagree eventually with Ori Belkind in his recent and important paper \cite{BELK}. I do not believe that Belkind's analysis exhausts the whole question, for the argument of the stepwise bigger spatial containers raises problems with respect to infinity (in fact it implies an argument against actual infinity), which I think is impossible to uphold with respect to Newton, since it would prevent him to make the cosmological leap which underpins the universality of the laws governing his system of the world. We saw before that, in order to formulate his First Law, Newton needs to posit the actual infinity of absolute space from the start. I do agree with Belkind, however, that, as long as measurable, finite physical systems are concerned, absolute space and time do not play any r\^{o}le except than as as warrant for the executability and repeatability of measurements. But with the following important caveat: the consistency of Newton's mechanics as a whole rests on the {\em a priori} equivalence of inertial and gravitational mass, encoded by his three laws, a condition which can be provided for only by taking the totality of the universe into account, and which finds its conceptual translation in the ideas of homogeneity and isotropy of absolute space and time. We find these latter ideas expressed explicitly in the famous Scholium to the Definitions:

\begin{quote} {\em I. Absolute, true and mathematical time, of itself and from its own nature, flows equally without regard to anything external.

II. Absolute space, in its own nature, without regard to anything external, remains always similar and immovable.} \cite[p. 13]{MOT}\cite[Var.,  p. 46 (18-28)]{VAR}.\end{quote}

\noindent Newton's two-layered approach allows him to be rigourous in principle and at the same time open to approximation whenever required to deal with real world problems. There is, in other words, no need for a complete causal, dynamical description of a given physical system on the relative level. {\em Otherwise, altogether no phenomenon could rightly be explained by its cause, unless, the cause of this cause and the cause of the prior cause were to be exposed and so successivley [and] continuously until the primary cause is arrived at.}\footnote{[CUL add. Ms. 9597.2.11: f.3$^r$] See for a discussion and more related material Ducheyne's paper on the General Scholium \cite{DUC}. The quote is in Ducheyne's translation.} This move mirrors a key aspect of Aristotle's criticism on Plato's causal theory.\footnote{It implies the reverse of the argument on the impossibility to attain by stepwise divisions an actual infinity of parts.} Newton chops off the embarassing infinities for all practical purposes --- relative motions --- but keeps them on the background as a warrant for consistency reasons. The glaring causal gap between the global and the local level is fillied up by the presence of an omnipresent God whose acts remain unfathomable to the finite human mind. {\em Because he actually and substantially exists in infinite space, God can act in and at every place} \cite[p. 95]{MGUIe}. The {\em Scholium Generale}, on the other hand, contains Newton's reflections on ``God, which is a relative word, and has a respect to servants, as distinct from the absolute terms of eternal, infinite or perfect as `titles which have no respect of servants'''. \cite[p. 131, p. 12]{STEW, SMET} The two-layered world picture clearly has a parallel in his theological conceptions.  God ultimately grounds the remarkable identity that Newton needs to give his mechanics a rock-solid logical, i.e., metaphysical, foundation. In a report destined for Leibniz, Fatio de Duillier gives  a clear summary of Newton's ideas:

\begin{quote} {\em  [concerning] Pag. 163 du Trait\'{e}  de Mr Hugens: Monsr Newton est encore indetermin\'{e} entre ces deux sentiments. Le premier que {\bf\em la cause de la pesanteur soit inherente dans la mati\`{e}re par une Loi imm\'{e}diate du Cr\'{e}ateur de l'Univers} et l'autre que la Pesanteur soit produite par la cause Mechanique que j'en ai trouv\'{e}e.}\footnote{Fatio \`{a} W. De Beyrie, pour Leibniz [1694], see Huygen's {\em Oeuvres Compl\`{e}tes} \cite[OH X, n$^{\circ}$2853, pp. 605-608]{HUY}. This is true beyond question, witness a manuscript in Newton's hand published by the Halls, the draft of a scholium on corr. 4 and 5 of prop. VI, Book III, where we read: {\em Huius autem generis Hypothesis est unica per quam gravitas explicari potest, eamque geometra ingeniosissimus D.N. Fatio primus excogitavit,} \cite[p. 313]{HAL}.}\end{quote}

\bigskip 

%\newpage

{\large \bf Leibniz}\\

\noindent According to a tenacious common place Leibniz, the philosopher, lost the battle against Newton, the scientist, for becoming the ``Founding Father of modern science''. The view is, basically, that Leibniz never succeeded to make the decisive step into natural science and remained steeped in a ``metaphysical'' attitude towards the questions at hand. The principal justification for the standard appraisal is mainly the content of the notorious Leibniz---Clarke correspondence \cite{ALEX} and certain XVIII-th century testimonia, often with an evident Newtonian bias. There are, however,  good reasons to think that this judgment merits reconsideration.\footnote{This topic is the central theme of a recent book by Daniel Garber, on what he calls the ``corporeal metaphysics'' of Leibniz's ``middle years''. I came across this book only when this paper was already largely finished, so that its influence remains limited. Evidently, my paper could only have gained from an earlier acquaintance. I nevertheless believe that my findings basically square with the main tenets of Garber's impressive work. See \cite{GARc}.}

Leibniz's dynamical ideas involve many subtleties, and moreover everything is interconnected with everything, which makes it impossible to deal with them appropriately within the scope of a single paper. But it is possible to filter out his stances on the topics relevant for our discussion without being unfair or overly simplifying. In the following, we shall therefore give only an outline of the relevant ideas on mechanical motion, the nature of matter, space and time, and the r\^{o}le and place of infinity in a specific period of his intellectual career. We shall focus on some texts from what Garber \cite{GARb} calls Leibniz's early and middle years (from around 1680 to 1700), with a special interest for the collection concerning ``la r\'{e}forme de la dynamique'' edited by Fichant, although with some occasional references to later texts as well.  Let us start by looking more closely to a relatively early text which is of considerable interest, {\em On the Nature of Body and the Laws of Motion}, written somewhere around 1680.\footnote{This text is in the collection provided by Ariew and Garber \cite[pp. 245-250]{GARa}. We use their translation. We do so for all English quotes from Leibniz, except when stated explicitly otherwise.}

\begin{quote} {\em There was a time when I believed that all the phenomena of motion could be explained on purely geometrical principles, assuming no metaphysical propositions, and that the laws of impact depend only on the composition of motions.\footnote{The reference is to an early work, the {\em Theoria motus abstracti}, from  1671. See for this whole period the discussion in Garber \cite{GARc}}.  But (...) I discovered that this is impossible, and I learned (...) that everything in nature can indeed be explained mechanically, but (...) that the principles of mechanics themselves depend on metaphysical (...) principles (...), that is, on contemplation of the most perfectly effectual, efficient and final cause, namely, {\em God} (...)}\end{quote}

\noindent At first glance this seems rather to confirm the reigning prejudice, but let us see what Leibniz has to say about how and why he changed his mind. The jump from `geometrical' to `mechanical' is permitted, according to Leibniz, when one assumes that everything is constituted by ``matter and its variations'' as the Epicureans held, which would imply the Cartesian-Newtonian conservation of motion. But the latter is precisely what he now thinks is impossible. Leibniz was led to this conclusion by his investigations into the laws governing the collision of bodies. In the text quoted above, he describes several simple collision experiments between solid bodies with equal or different mass, and with equal or different speeds, on a fixed or on a moving (boat) surface.  

\begin{quote} {\em I sought a demonstration for this [the behaviour of two bodies after collision] from my assumption that, in body, nothing can be considered except bulk [{\em moles}], that is, extension and impenetrability, or what comes down to the same thing, the filling of space or place. Moreover, I assumed that nothing could be considered in motion except (...) the change of place. But if we want to assert only what follows from these notions, we will say that the reason [{\em causa}] why a body impels another must be sought in the nature of impenetrability (...)} \cite[p. 246]{GARa}\footnote{I.e., composed of impenetrable atoms, as is clear from his reference to the ``Epicureans''. The fundamental ``elasticity'' of matter is an ongoing theme with Leibniz, and a hotly debated issue in the correspondence between Leibniz and Huygens.}\end{quote}

\noindent The reader will appreciate the similarity between some aspects of this argument and the style of reasoning in Newton's {\em De Gravitatione}. But the purely mechanical argument leads to results which contradict experience says Leibniz, on the basis of experiments which he might have done himself. On the given assumptions, if two solid bodies of equal weight would bounce into each other with equal speed along the same line and from opposing directions, they both should be at rest after the collision. It they collide into each other with different speeds, then the {\em the slower body $B$ is carried of by the faster body $A$ (...) with [a] speed which is the difference between the prior speeds} \cite[p. 246]{GARa}. But this is not the case. This failure  is, according to Leibniz, related to the material constitution of bodies --- there are no truly solid bodies; every body, however hard, has some elasticity\footnote{Cfr. \cite[p. 42, p. 53]{FICH}. In the {\em Specimen Dynamicum} [1691],  Leibniz will connect this idea to the existence of the aether, and thus to the cause of gravity: {\em no body is so small that it is without elasticity, and furthermore, each body is permeated by a fluid even subtler than it is. And thus, {\em there are no elements of bodies}, nor is there maximally fluid matter, nor are there little solid globes (unintelligible to me) (...) Rather, the analysis proceeds to infinity.} \cite[pp. 132-133]{GARa}.} --- and because the argument leaves out of consideration the effects of what we would now call inertia: {\em For to say that matter resists motion, and that the whole composed of A and B together now moves more slowly than A did before, is to claim that there is something that cannot be derived from the simple nature of body and motion of the sort we assumed above, if in that nature we understand nothing but the filling and change of place} \cite[p. 247]{GARa}. Leibniz follows here a pattern of argumentation laid out at first by Descartes, and basically retained by everybody who afterwards contributed anything to the development of XVII-th science of motion.\footnote{Even though Descartes's own laws of motion proved wrong (as he himself already suspected), and were corrected afterwards by Huygens, Wren, Wallis and Mariotte, though their solutions remained incomplete. A good overview of the history is in the introduction by Fichant \cite[p. 15 sq.]{FICH}.} His solution is to suppose that in bodies reside certain immaterial powers that regulate the proportions between speed and magnitude. It are these powers, and not motion, which are conserved in the world. In a remarkable text of january 1678, {\em De corporum concursu}, Leibniz came to the formulation of a more ``abstract''  framework for the description of bodies in motion which contains already all the elements of the mature dynamics: the relativity of (local) motion, his definition of force as the ``quantity of the effect'', determined by the height which can be attained by a moving body --- i.e., its potential energy --- as well as the principle of conservation of what he will later call {\em vis viva}, of $mv^2$ instead of $mv$. This text has been edited for the first time by Fichant in 1994. Fichant summarises it in his introduction as follows:

\begin{quote} {\em Leibniz a ainsi caract\'{e}ris\'{e} comme ``r\'{e}forme'' {\em (reformatio)} la nouvelle formulation de d\'{e}finitions et de principes qui rendent possible une mise en \'{e}quation coh\'{e}rente et compl\`{e}te des r\`{e}gles du mouvement pour le probl\`{e}me \'{e}l\'{e}mentaire du choc direct de deux corps et dans tous les cas de figure sous lesquels il peut \^{e}tre particularis\'{e}. L'acte essentiel en est la red\'{e}finition de la force par la mesure de son effet, associ\'{e}e \`{a} la substitution du carr\'{e} de la vitesse \`{a} la vitesse simple dans son expression m\'{e}trique, figurant dans {\bf \em un principe de conservation} dont le domaine de validit\'{e} contingente est nomm\'{e} par Leibniz {\bf \em ``syst\`{e}me''}.}\cite[p. 15 (our bold)]{FICH}.\end{quote}

\noindent The essential move here is from local motion to global conservation, from mechanics to metamechanics --- as Leibniz indicates himself in the earlier quote. The metaphysical principle involved is that {\em the global effect always equals its full cause}.\footnote{{\em Quod effectus integer sit semper aequalis causae suae plenae}. \cite[p. 50, ft. 2 sq.]{FICH} (My translation). This comes back in the {\em Specimen Dynamicum} \cite[p. 129]{GARa}. Cfr. {\em Effectus integer aequipollet causae plenae}, Quoted by Y. Belaval \cite[p. 129]{BELA}.} It is relevant to stress that Leibniz takes recourse to this only after it appeared again and again that a satisfactory general solution to the problem of collision was impossible on the basis of mechanical quantities of solid bodies alone. These theories all violate the principle of invariance of the common centre of gravity, a formulation of the relativity principle Leibniz inherited from Huygens: if the centre of gravity of two colliding bodies differs from their point of collision, imagine them sitting on a steadely floating boat that covers the required distance between the two points in the appropriate time, and {\em for any observer on the riverbank}, the symmetry will be restored.\footnote{This is already in Huygens's {\em De motu corporum ex percussione} [1656]. Whence Leibniz calls it ``la m\'{e}thode du bateau''. Cfr. \cite[p. 14, 31, 190 sq.]{FICH}} This is the germ of Leibniz's ideas of ``system'' and ``full dynamical explanation'' that will play such an important r\^{o}le in his later work. It also explains while Leibniz insists on the fact that forces, dynamical invariants, are {\em real}, while motions are appearances subject to phenomenal relativity (``real'' to the mind only) \cite[p. 190.]{FICH}.

The {\em concursu} is the eloquent witness to Leibniz's  efforts to ``get finally out of this labyrinth'' \cite[p. 50]{FICH}, both theoretically and experimentally. It contains tables with detailed measurement outcomes of experiments already envisaged in 1677\footnote{See the Letter to Jean Berthet, {\em S\"{a}mtliche Schriften und Briefe} [Akademienausgabe], II, vol. 1, p. 383.}, and pages and pages of calculations based on them. To account for phenomena of motion, one has not only to consider their mere mechanics, but also the forces originating from the internal constitution of the bodies involved, and the influence of the global ``system'' of which they are part. What one needs is a theory that takes at the same time the mechanical, the internal, and the inertial aspect of motion into account. The key idea is that no body is absolutely hard or solid, so that the ``repercussion'' after impact which we observe in most of the cases can be explained by the forces arising from their internal elasticity. This universal elasticity assumption, together with the conservation and relativity principles already mentioned, is the core of the theory Leibniz ventures into succesfully towards the end of the {\em De corporum concursu}, after his earlier failed attempts based on the common mechanical (Cartesian) assumptions.\footnote{Again, this is the origin of his lifelong rejection of atomism, and one of the few fundamental disagreements that continues to surface in his correspondence with Huygens, an important debate to which we shall come back in a another article. Leibniz reiterates this point on several occasions, cfr. e.g. the quote from the {\em Specimen Dynamicum} above. \cite[pp. 132-133, p. 136]{GARa}.} 

The great advantage of this approach is that every quantity involved relates directly to some observable characteristic of motion in the sensible world, while at the same time the global viewpoint of the ``system'' arises in a natural way as a crucial theoretical device indispensible to ``save the phenomena''. The more encompassing level of ``system'' is not reducible anymore to the mere mechanics involver: {\em to ground the reality of motion, Leibniz tirns away from the purely geometrical, and to the underlying causeof change.} \cite[p. 111]{GARc} The effect of the internal constitution of bodies, as well as the phenomenon of resistance, necessitate this move. {\em (...) resistance is, itself, a kind of activity.} \cite[p. 117]{GARc} But he would rather not ascribe these features to a direct intervention of God, as Newton did. Instead, he hoses to add these powers to the physical description of the body as a whole:

\begin{quote} {\em Leibniz's preferred solution is to ground force and activity directly in body itself. That is, to inert matter we must add ``powers'' or forces, that ``by which speed is adjusted to magnitude.'' (...) forces or powers are identified with the forms that Leibniz wants to attribute to bodies: {\bf \em  forms just {\em are} powers or forces}. And if forms are understood in this way, then adding force and activity to body {\em is just} to add form.} \cite[p. 118 (our bold)]{GARc}.\end{quote}

\noindent This feature is what marks out Leibniz's theory as metaphysical in the proper sense, as he himself realises very clearly. This separation of the local from the global also paves the way for the sound application of his definition of existence: to exist means to be free from contradictions; since the phenomenal reality of motion does in no way threaten the internal consistency of substantial things. Leibniz with some reason calls this new approach a ``reform'' ({\em reformatio}) in the science of motion, thus indicating the originality and the importance he accorded to his own achievement, years before the metaphysical synthesis of the {\em Discours de M\'{e}taphysique} was conceived. So Leibniz had a solid basis for his own dynamics long before the publication of the {\em Principia}, completing, so to say, the corrected version of Cartesian mechanics that had been devised by Huygens in 1656, 1669 and 1673. In the {\em Specimen Dynamicum}, a much later text which he published after reading Newton's {\em Principia}, Leibniz writes: {\em That is, we acknowlegde that all corporeal phenomena can be derived from efficient and mechanical causes, but we understand that these very mechanical laws as a whole are derived from higher reasons} \cite[p. 126]{GARa}.  It thus seems to be as Gerhardt states in his edition of the philosophical writings, that Leibniz's dynamical key concepts will inform the whole of his metaphysics, rather  than the other way around [GP III, p. 48, quoted in \cite[p. 9]{FICH}]. Only later on Leibniz will make the reverse move and subordinate dynamics to metaphysics, ``which treats of cause and effect'' \cite[p. 252]{GARa}.

Now, how does all this relate to the discussion on central forces and the cause of gravity? Leibniz, following Huygens and Descartes, believed that gravity was caused by centrifugal forces working on bodies in a rotating fluid medium, the {\em aether}.\footnote{{\em the repercussion and bursting apart [of a body after impact] arises from the elasticity it contains, that is, from the motion of the fluid aetherial matter permeating it, and thus it arises from an internal force or a force existing within itself.} \cite[p. 135]{GARa}} Centrifugal forces themselves are the result of circular motion. Huygens had introduced his centrifugal force in order to be able to quantify the force to which a body in circular motion about a centre is subjected by equating it to the tension in the string holding its weight as a bob in a pendulum clock modelling the system  under consideration, another brilliant move which proved crucial to many later developments.\footnote{An excellent study of Huygens's approach is J. Yoder, {\em Unrolling time: Christiaan Huygens and the mathematization of nature}, Cambridge University Press, Cambridge, 1988.} Leibniz inherits this idea:

\begin{quote} {\em For if we assume something we call solid is rotating around its center, its parts will try to fly off on the tangent; indeed, they will actually begin to fly off. But since this mutual separation disturbs the motion of the surrounding bodies, they are repelled back, that is, thrust back together again, as if the center contained a magnetic force for attracting them, or as if the parts themselves contained a centripetal force. Thus, the rotation arises from the composition of the rectilinear nisus for receding on the tangent and the centripetal conatus among its parts.} \cite[pp. 135-136]{GARa}\end{quote}

\noindent The non-existence of truly solid (absolutely hard) bodies functions as a key stone to the whole of the complex building which is Leibnizian dynamics, for it also shaped his ideas on the nature of planetary motion. It is clear that Leibniz subscribes to the plenum-ontology implied in the vortex-theory. But he nevertheless admits that there are other possible hypotheses that explain the experimentally available data equally well: motion {\em in vacuo} with inertia and gravitation, or the antagonistic tendencies working on a body in circular and radial motion in a fluid medium.\cite[p. 32]{MELb}  But even though there are equivalent hypotheses that could `save the phenomena' of celestial mechanics,  the elasticity problem shows why Newton's bucket argument (and thus his claims with respect to the truth of his ``system of the world'') does not hold: one can never prove that it covers all relevant factors generating the observed phenomena, which is required for a fully ``systematic explanation of things''\footnote{{\em whenever we are dealing with the equivalence of hypotheses, we must take into account everything relevant to the phenomena.} \cite[p. 137]{GARa}}: 

\begin{quote} {\em  (...) the rotation of a solid body requires an account of solidity, and it may be that solidity (or fluidity) arises from interactions between the solid body and its surroundings. Thus, unless we take into account {\bf \em the full dynamic explanation} that tells us what makes a body solid and how its motion is determined relative to the surroundings, it is not possible to take the inertial effects produced by rotation as a phenomenal criterion of true motion.} \cite[p. 48 (our bold)]{BELKb}\end{quote}

\noindent Leibniz moreover holds that choice between hypotheses amounts to selecting ``the simplest hypothesis most suitable for explaining the phenomena'' \cite[p. 135, ft. 173]{GARa}. {\em Ideally}, the simplest is also the best. We need to stress once more the crucial difference between substantial things and the phenomenal relations between them. This brings us immediately to Leibniz's conception of space. Space is phenomenal, as he repeates over and over again in his correspondence with Clarke, not because of some {\em a priori} reason, but because space, the totality of place, is an abstraction following from the  relativity of position, and place is just an equivalence class of possible positions, in contradistinction to the {\em relations of situation} of real objects: the first is not a subject nor an attribute to it, so that it cannot be anything else than an ``ideal thing'', but the latter is. This difference can be captured by means of an analogy: it is not possible for {\em me} to be {\em you}, but I can perfectly well occupy your {\em place}, as it is related to a third, external observer. The concrete situation of bodies with respect to each other, however, is not: no-body can step outside itself to try to look at itself externally:

\begin{quote} {\em though the places of the three particles may be the same in each case, their {\em relations of situation} are not; since the latter are `affections' of the bodies at particular moments of their histories, there is a genuine difference in the two cases [in which two bodies swapped places], {\em viz.} as part of the respective monadic perceptual histories. (...) {\bf \em Extrapolating this actual relation into a possible relation for all bodies whatever}, yields an ideal system of possible and actual relations which is conceived as extrinsic to bodies; this is what we call place, and {\em space}.} \cite[p. 204 (my bold).]{WIN}\end{quote}

\noindent By hypostasing place into {\em absolute} space, one falls into the trap of conflating the ideal with the real, the species with the individual, so to say. A.T. Winterbourne, in his ground-breaking paper on Leibniz's conceptions of space, insists again and again on the fact that conflating the different levels of Leibniz's system leads into fundamental misunderstandings. We cannot do better than let Leibniz explain himself once again: {\em For even though force is something real and absolute, motion belongs among phenomena and relations, and we must seek truth not so much in the phenomena as in their causes.} \cite[p. 131]{GARa}  Even experiment will not help you out, because in the end, to measure means to relate something to yourself, so that all measurement presupposes an irreducible point of view, which makes the observed relation different depending on the perspective taken on the observed entities \cite[p. 205]{WIN}.\\

\noindent We already indicated that, to Leibniz, the constitution of matter and the nature of infinity are also related to one another, as a consequence of the universal elasticity criterion: {\em no body is so small that it is without elasticity, and furthermore, each body is permeated by a fluid even subtler than it is. And thus, {\em there are no elements of bodies} (...)  nor are there little solid globes (unintelligible to me) (...) Rather, {\bf \em the analysis proceeds to infinity}.} \cite[pp. 132-133 (my bold)]{GARa} To say that there are perfectly impenetrable solid bodies is to say something incomprehensible, which required a direct action of God's free will into the world; as we saw, this position implies a direct criticism of Newton's underlying metaphysics.\footnote{Cfr. Leibniz in the {\em Specimen Dynamicum}: {\em I believe that there is no natural truth in things whose explanation [{\em ratio}] ought to be sought directly from divine action or will.} \cite[p. 125]{GARa}.} It is not space that grants existence to things; it is the fact that they are present consistently in all their aspects in the Divine mind \cite[p. 229]{KNE}. Thus the problems arsing from the absoluteness of space and those related to infinity are closely interconnected.\footnote{Cfr. A. Lamarra \cite[p. 189]{LAMA}. Leibniz's position with respect to infinities in mathematics has been studied throroughly in recent papers by Richard Arthur, see, e.g., \cite{ART}.}

But since in God the possible and the actual, the real and the ideal coincide, for Him the infinities do exist. Even if {\em we} do not have any direct  access to the ``infini incomparable'', it still exists before God's eye, and we can have, thanks to the inherent rationality of creation and the use of a correct (mathematical) method for invention, a certain access to it in the ideal realm. All possible worlds are  {\em simultaneously} present before God's all seeing eye. One of these possibilities --- the best --- is our world. In this context, it makes perfect sense to understand the later Leibniz's Principle of Sufficient Reason as the full metaphysical translation of his earlier, dynamical simplicity requirement. The level of mechanics is the local, actual universe, our world, of which we assume that it fulfllls the requirement of finitude. This, however, does by no means apply to the level of ``absolutely everything'', which is  God's perspective only, and which coincides with the infinitely large `space of possibilities'.\footnote{Leibniz therefore breaks with another basic tenet of Cartesian mechanical metaphysics: that only the indefinite is real. \cite[pp. 275-276]{BELAb}} 

\begin{quote} {\em 225. The wisdom of God, not content with embracing all the possibles, penetrates them, compares them, weighs them one against the other (...) It goes even beyond the finite combinations, it makes of them an infinity of infinites, that is to say, an infinity of possible sequences of the universe, each of which contains an infinity of creatures. (...) The result of all these comparisons and deliberations is the choice of the best from among all these possible systems (...).  Moreover,  {\bf \em all these operations of the divine understanding, (...) always take place together}, no priority of time existing among them.} \cite[\S 225]{LEIB}; cfr. \cite[p. 242]{BELA}.\end{quote}

\noindent The Newtonian difference --- ontological and epistemological  --- between the relative and the absolute  becomes with Leibniz the difference between the possible and the actual. Eventually, for the mature Leibniz, the world of possibilities determines the perspective and thus the modus of existence of every concrete being. This presence of an ideal, i.e. formal, level even from the individual point of view is commensurable with his notion of system, and leads to his later conception of {\em monad}\footnote{{\em Car Dieu tournant pour ainsi dire de tous c\^{o}t\'{e}s et de toutes les fa\c{c}ons le syst\`{e}me g\'{e}n\'{e}ral des ph\'{e}nom\`{e}nes (...) et regardant toutes les faces du monde de toutes les mani\`{e}res possibles, puisqu'il n'y a point de rapport qui \'{e}chappe \`{a} son omniscience, le r\'{e}sultat de chaque vue de l'univers, comme regard\'{e} d'un certain endroit, est une substance qui exprime l'univers conform\'{e}ment \`{a} cette vue (...)} \cite[II, p. 95]{GERH}}, the endpoint of his earlier idea to think of forces as forms.

In Leibniz's metaphysics, no direct interference of divine will with the course of events is needed; everything is mechanically and metaphysically transparent once it is placed in its appropriate context. But ultimately, Leibniz does not get away from the original paradoxes either, because a God seeing all possibilities at once leads into inconsistency. This can be seen easely if one recasts Leibniz's basic example of the possible {\em relations of situation} of objects  into elementary counterfactual propositions. ``Counterfactual'' is taken here in the straightforward sense of ``if -- then'' statements conform to the basic data concerning a given world. Observations are made always from within the perspective of one ``inhabitant'' only. The question is then whether one could define the possible outcomes of the (context-dependent!) situation-measurements {\em globally}, while the measurements can be executed only {\em locally}. It has been shwon by specker in a famous argument that this is inconsistent \cite{SPE}. Only in a strictly context-independent case an overall evaluation is consistently possible. Specker, in his original 1960 paper, makes the link to the problem of an omniscient God himself:

\begin{quote} {\em In a certain sense, however, these issues were anticipated by scholastic speculations concerning ``infuturabilien'', [future contigencies --- transl.], i.e., the question of whether God's omniscience includes events which would occur if something were to happen which in fact does not happen.}\footnote{On the basis of the arguments in this paper, Kochen and Specker will develop their famous theorem  for quantum mechanics \cite{KOCH}. The original paper, however, deals with the general case of undecidable propositions independently of QM. For the quote: \cite[pp. 239-246]{SPE}. Translation in C.A. Hooker, see p. 138.}\end{quote}

\noindent In his comment, Svozil comments dryly: ``Today, the scholastic term ``infuturability'' would be called ``counterfactual'' \cite[p. 79 sq.]{SVO}.  Remember that already Aristotle had to introduce the conservation of identity through time  --- his principle of the excluded middle --- to allow for his logic to treat {\em contigentia futura} consistently \cite{FRE}. Moreover, whether one understands Leibniz as saying that for God everything is actual, or rather that all possibilities are modally present in God's mind does not make a difference, for in both cases the argument remains applicable \cite{ROND}.

\subsection*{\sc Conclusion}

In this contribution, we tried to shed some light on the inconsistencies that arise in theories that attempt to describe or explain the world at large by looking at the way they deal with the problem of causality. All metaphysical theories encompass some notion of causality. Even in the absence of a common notion of causality, there appears to be a common strcutural framework exhibited by all theories that use the concept, implicitly or explicitly. Meyerson showed that, in order to use any notion of causality consistently, one has to assume the {\em conservation of identity through time}. It is possible to trace this principle back to classical metaphysics, where it functions as the device to defeat the ancient paradoxes of plurality and motion, because it allows for the separation between the global (universal) and the local level within a given picture of the world. We therefore identify this conservation as {\em the} fundamental metaphysical principle: any theory calling upon it is metaphysical in a very rigorous sense, defined on the structural level of the theories concerned. When causality coincides with lawful behaviour, as is the case in modern scientific theories, one gets a very specific instance of this principle. Lawfulness is the cloak for the underlying metaphysics in early modern science and in science per se. Using these insights, we have identified and compared metaphysical theories by means of their common structural characteristics, rather than by explicit ontological content. More specificially, we proposed a comparison between the underlying metaphysics in the dynamical theories of Newton and Leibniz. Newton makes the distinction between levels in a very explicit way, but pays this with a complete lack of transparency when it comes to the causal mechanism, for which he ultimately has to call upon direct interference by God. His God thus exhibits the ancient paradoxes in a way similar to Aristotle's Prime and Unmoved Mover. Leibniz on the other hand has a much more transparent dynamical theory with global and local levels, mechanical causes and conservation principles, but he runs into trouble where the relation between God and His creation is involved: the paradox of an omniscient observer. This, as has been shown by Specker in a famous argument, is inconsistent in its own right.

\end{document}